\documentclass[12pt]{spieman}  
\usepackage{amsmath,amsfonts,amssymb}
\usepackage{color}
\usepackage[colorlinks=true, allcolors=blue]{hyperref}
\usepackage{comment}
\usepackage{here}
\usepackage{siunitx}
\usepackage{multirow}
\usepackage{bm}
\usepackage{orcidlink}
\usepackage{academicons}
\usepackage{lscape}
\usepackage{longtable}
\usepackage{threeparttable}
\usepackage{graphicx}
\usepackage{setspace}
\usepackage{tocloft}

\title{Optimization of x-ray event screening using ground and in-orbit data for the Resolve instrument onboard the XRISM satellite}

\author[a,b,*]{Yuto Mochizuki}
\author[b]{Masahiro Tsujimoto}
\author[c]{Caroline A. Kilbourne}
\author[d]{Megan E. Eckart}
\author[e]{Yoshitaka Ishisaki}
\author[b]{Yoshiaki Kanemaru}
\author[c]{Maurice A. Leutenegger}
\author[f]{Misaki Mizumoto}
\author[c]{Frederick S. Porter}
\author[g]{Kosuke Sato}
\author[h]{Makoto Sawada}
\author[h]{Shinya Yamada}

\affil[a]{Graduate School of Science, The University of Tokyo, Bunkyo-ku, Tokyo, 113-0033 Japan}
\affil[b]{Institute of Space and Astronautical Science, Japan Aerospace Exploration Agency, Sagamihara, Kanagawa, 252-5210 Japan}
\affil[c]{National Aeronautics and Space Administration, Goddard Space Flight Center, Greenbelt, MD 20771, USA}
\affil[d]{Lawrence Livermore National Laboratory, Livermore, CA 94550, USA}
\affil[e]{Department of Physics, Tokyo Metropolitan University, Hachioji, Tokyo 192-0397, Japan}
\affil[f]{Science Education Research Unit, University of Teacher Education Fukuoka, Munakata, Fukuoka 811-4192, Japan}
\affil[g]{International Center for Quantum-field Measurement Systems for Studies of the Universe and Particles, Tsukuba, Ibaraki 305-0801, Japan}
\affil[h]{Department of Physics, Rikkyo University, Toshima-ku, Tokyo 171-8501, Japan}

\begin{document}
\maketitle

\begin{abstract}
 The XRISM (X-Ray Imaging and Spectroscopy Mission) satellite was successfully launched
 and put into a low-Earth orbit on September 6, 2023 (UT). The Resolve instrument
 onboard XRISM hosts an x-ray microcalorimeter detector, which was designed to achieve a
 high-resolution ($\leq$7~eV FWHM at 6~keV), high-throughput, and non-dispersive
 spectroscopy over a wide energy range. It also excels in a low background with a
 requirement of $< 2 \times 10^{-3}$~s$^{-1}$~keV$^{-1}$ (0.3--12.0~keV), which is
 equivalent to only one background event per spectral bin per 100~ks exposure. Event
 screening to discriminate x-ray events from background is a key to meeting the
 requirement.
 We present the result of the Resolve event screening using data sets recorded on
 the ground and in orbit based on the heritage of the preceding x-ray
 microcalorimeter missions, in particular, the Soft X-ray Spectrometer (SXS) onboard
 ASTRO-H. We optimize and evaluate 19 screening items of three types based on (1) the
 event pulse shape, (2) relative arrival times among multiple events, and (3) good time
 intervals. We show that the initial screening, which is applied for science data
 products in the performance verification phase, reduces the background rate to $1.8
 \times 10^{-3}$~s$^{-1}$~keV$^{-1}$ meeting the requirement. We further evaluate the
 additional screening utilizing the correlation among some pulse shape properties of
 x-ray events and show that it further reduces the background rate particularly in the
 $<$2~keV band. Over 0.3--12~keV, the background rate becomes $1.0 \times
 10^{-3}$~s$^{-1}$~keV$^{-1}$.
\end{abstract}

{\noindent \footnotesize\textbf{*}Yuto Mochizuki,  \linkable{mochizuki@ac.jaxa.jp} }

\begin{spacing}{1}
\thispagestyle{plain}
\pagestyle{plain}

\section{Introduction}\label{sec:intro}
The X-Ray Imaging and Spectroscopy Mission (XRISM)
\cite{tashiro2024} was successfully launched on September 6, 2023 (UT) into a low-Earth
orbit of an inclination of 31 degrees from JAXA's Tanegashima Space Center using an
H-IIA rocket. One of the scientific instruments is Resolve
\cite{ishisakiStatusResolveInstrument2022}, which was designed to achieve
high-resolution ($\sim$7~eV FWHM at 6~keV), high-throughput, non-dispersive spectroscopy
over a wide energy range (0.3--12~keV) based on x-ray microcalorimetry. The energy
resolution of $\sim$4.5~eV at 6~keV was actually achieved in orbit\cite{porter2025}.
Resolve was developed under an international collaboration between JAXA and NASA, with
participation from ESA based on the heritages of the X-Ray Spectrometer (XRS) onboard
the ASTRO-E satellite\cite{porterAstroE2XraySpectrometer2004}, XRS2 onboard the the
ASTRO-E2 satellite\cite{kelleySuzakuHighResolution2007}, and the Soft X-ray
Spectrometer (SXS) onboard the ASTRO-H satellite\cite{kelleyAstroHHighResolution2016}.

The Resolve x-ray microcalorimeter spectrometer excels not only in energy resolution but
also in low background. The requirement for the background rate is $< 2 \times
10^{-3}$~s$^{-1}$~keV$^{-1}$, which is equivalent to only one background event per
spectral bin per 100~ks exposure. A background level significantly lower than this
requirement was indeed achieved in orbit with the SXS
\cite{kilbourneInflightCalibrationHitomi2018}, which has almost the same design with
Resolve\cite{ishisakiStatusResolveInstrument2022}. A combination of hardware and
software was needed.
For the hardware, the anti-coincidence detector
\cite{kilbourneDesignImplementationPerformance2018} worked efficiently to reduce the
background rate to 23\% by
providing a signal that software screening used to remove events caused by cosmic-ray hits
\cite{kilbourneInflightCalibrationHitomi2018}.
For the SXS software,
in addition to the screening based on the anti-coincidence detector signals, a set of event-screening criteria was developed to discriminate signal events by cosmic x-ray photons from other events of various origins, which we collectively call background events.
In total, the combination of the two
reduced the SXS background rate to 7.3\% of the original rate \cite{kilbourneInflightCalibrationHitomi2018}.
Owing to this low background,
a new scientific discovery was made only with 17 events from a supernova remnant N132D
with SXS\cite{hitomicollabo}.

\medskip

The purpose of this paper is to describe the event screening of Resolve and present its
optimization and performance using the in-orbit and ground data sets.  We follow the
screening developed for the SXS\cite{kilbourneInflightCalibrationHitomi2018}, but some
new screening items and algorithms were developed for Resolve. We made assessments of
the existing and new items using the data taken with Resolve. Satellite-level screening,
such as telemetry loss, availability of star-tracking control, and angular distance from
targets, are out of scope for this paper. The methods presented in this paper are the
latest at the time of the performance verification phase. Updates in the future will be
shared with the community through the documents provided for guest observers.

A larger volume of data is available for event screening in Resolve than in SXS, both in
orbit and on the ground. As for the in-orbit data, SXS was lost unexpectedly early only
34 days after launch \cite{tsujimotoInorbitOperationSoft2017} due to the loss of the
spacecraft attitude control in 2016, yielding only 366~ks integration of good background
data constructed from Earth occultation times. Resolve has been operated in orbit under
50~mK since October 9, 2023, and longer integration by an order has already been
accumulated \cite{maeda2025}.
%

As for the ground data, Resolve acquired a more complete set of data at the highest
level of integration\cite{tashiro2024,eckart2025,mizumotoHighCountRate2025} compared to SXS \cite{eckartGroundCalibrationAstroH2018},
including data in the gate valve (GV) open configuration for low-energy calibration
(below 4~keV), a more complete set of exposures in the extended high energy band (up to
25~keV), and exposures with monochromatic x-ray lines at additional line energies (six
lines from 4.5--11.9~keV versus two lines from 5.4--8.0~keV). Here, the GV is an
aperture door with an x-ray transmissive window
\cite{midookaXrayTransmissionCalibration2021} to keep the cryostat leak-tight before the
launch. A special apparatus was used to open the GV on the ground for low-energy data
acquisition.
Such data sets are not (yet) available in the in-orbit. Monochromatic x-ray sources and
high energy lines above 12~keV are unavailable among celestial sources with sufficiently
high fluxes.
Data have not yet been taken with the GV open in orbit, as the initial
attempts to open it were unsuccessful\cite{tashiro2024}.
These ground data sets are thus unique for developing algorithms for
event screening.

\medskip

The structure of this paper is as follows. In \S~\ref{sec:inst}, we provide a
brief description of the Resolve instrument to make the paper self-explanatory as much
as possible. We focus on the signal processing both in orbit and on the
ground. References for the other aspects of the instrument can be found in a review
\cite{satoHitomiXRISMMicrocalorimeter2023} and other contributions in this volume. In
\S~\ref{sec:individual}, we explain the individual event screening items with a
particular emphasis on new assessments made with Resolve using various ground and
in-orbit data. Software implementations for these screening items are available online
at \url{https://heasarc.gsfc.nasa.gov/lheasoft/help/rslpipeline.html}.  In
\S~\ref{sec:initial}, we present the performance of the initial screening, which is used
for data products in the performance verification phase, to meet the requirement using
the in-orbit data. In \S~\ref{sec:additional}, we discuss further additional screening
to increase the signal-to-noise ratio to further extend the Resolve use cases toward
faint sources. The paper is concluded in \S~\ref{sec:conclusion}.

\section{Signal processing}\label{sec:inst}
\subsection{Detector and analog signal processing in orbit}\label{s2-1}
The x-ray microcalorimeter detector consists of $6 \times 6 = 36$ pixels (pixel 0--35)
constituting an array with a side length of 5~mm.  The array is situated in the center
of a Si frame that fans out the electrical connections to the pixels. Portions of the
front and back of the frame are coated with Au to assist in the transport of heat,
including, on the back, the portions of the frame between
pixels\cite{kilbourneDesignImplementationPerformance2018}. One of the read-out channels
is connected to a pixel (pixel 12) that is displaced from the array and outside of the
aperture and is under constant illumination by a dedicated $^{55}$Fe calibration source.
One corner pixel of the main array is not read out to acommodate this calibration pixel.
Each pixel comprises a HgTe x-ray absorber and a suspended Si thermistor. The energy
deposited by photoelectric absorption of x-rays is dissipated into heat, which is
conducted to the 50~mK heat sink with a thermal time scale of a few ms
\cite{chiao2018}.

Underneath the substrate of the x-ray microcalorimeter, a Si detector is placed with a
side length of 10 mm and a thickness of 0.5 mm that is used as an anti-coincidence
detector (anti-co)\cite{kilbourneDesignImplementationPerformance2018}.
Cosmic-ray particles (mostly minimum-ionizing protons) interact with the detectors mainly via ionization losses along their track lengths and are detected both by the anti-co (as charge) and the microcalorimeter (as heat).  The anti-co spectrum peaks at 200~keV, while the spectrum of calorimeter events coincident with anti-co events peaks at 6~keV, with these peaks corresponding to the energy deposited by a relativistic proton at normal incidence to the anti-co (0.5~mm thick Si) \cite{eidelmanReviewParticlePhysics2004} and a calorimeter absorber (0.0085~mm thick HgTe), respectively.  The pulse characteristics of cosmic-ray and X-ray events in the calorimeter are identical, which is why the anti-co is essential for rejection of particle events.
The anti-co detector is read out by two readout circuits for redundancy.

Each of the 36 microcalorimeter pixels and two anti-co channels is read out individually using
a pair of signal and return lines without multiplexing.  The output impedance of each
detector circuit is reduced by a Junction Field Effect Transistor (JFET) source-follower
operated at 130~K\cite{chiao2018} before being transferred
from inside to outside of the cryostat via a feed-through.
The analog signal is received by the analog electronics called X-ray box (XBOX)
\cite{chiao2018} placed outside of the cryostat. The XBOX provides the bias voltages to
the microcalorimeter and anti-co detectors and processes their output by shaping,
amplifying, filtering, and sampling at a rate of 80 $\mu$s (12.5 kHz) with a bipolar 14
bit depth analog-to-digital converter (ADC). The signal-chain hardware up to this part
is provided by NASA.

\subsection{Digital signal processing in orbit}\label{s2-2}
The rate of the entire time-series data (12.5 kHz $\times$ 16 bit $\times$ 38 channels)
is too large to fit into the telemetry bandpass to ground stations. Therefore, a
significant data reduction needs to be made in the orbit. Events are detected and their
time-series data (we call ``pulse'' hereafter) are processed. A handful of
characteristic values are derived for each event, which are downlinked at five
$\sim$10-minute contacts per day\cite{boyceDesignPerformanceASTROE1999}. When resources
are available in the data recorder and the telemetry bandpass, the pulse data of
randomly selected events are downlinked for diagnostic purposes, which are called pulse
records. This data reduction is performed using the onboard digital electronics called
the Pulse Shape Processor (PSP) provided by
JAXA\cite{ishisakiInflightPerformancePulseprocessing2018}.

The digitized time-series data are sent from the XBOX to the PSP, which consists of two
identical units (PSP-A and PSP-B). Each unit hosts one Field Programmable Gate Array
(FPGA) board and two Central Processing Unit (CPU) boards. One FPGA board processes 18
microcalorimeter and 1 anti-co channel. One CPU board processes 9 microcalorimeter and
1 or 0 anti-co channel by default. The two anti-co channels are readouts of the same
set of events. We used data from the A side in the presented analysis. All cross-channel
processing, such as vetoes, is done on the ground.

The FPGA board is responsible for calculating the time derivative
of the incoming microcalorimeter data (ADC samples) from the XBOX using a box car filter.
It triggers event
candidates based on a threshold applied to the time derivative. The threshold was changed
once in orbit (at 19:50 on October 31, 2023) from 75 to 120 as a precaution before the
first attempt to open the GV and has been left at that value. The FPGA board
also detects events in the anti-co
detector data by applying a threshold in ADC samples and derives characteristic values of
individual pulses: (1) the arrival time when the threshold is exceeded, (2) the maximum
ADC sample value (\texttt{PHA}), and (3) the number of consecutive intervals between
samples exceeding the threshold (\texttt{DURATION}). We recognize events with a minimum
\texttt{PHA} of 71 (equivalent to an energy deposition of 30.5 keV at PSP A side) and a minimum
\texttt{DURATION} of 2 to be cosmic-ray events.

The CPU board is responsible for detecting secondary pulses overlapping others and
deriving characteristic values of each pulse in the microcalorimeter data. One of the
values is the event grade, which is based on how an event is isolated in time from the
others on the same channel. Events with no preceding pulse in a fixed time are called
p(rimary) and the others s(econdary). Events without others within $\pm$69.92~ms are
H(igh) grade, those without others within $\pm$17.52~ms are M(id) grade, and the rest
L(ow) grade. For the H and M grade events, the time series data of a pulse is correlated
with templates for better characterization of the energy and arrival time.  The template
is made for each pixel from the average pulse shape of numerous events from 5.9~keV
x-ray photons and noise spectra obtained in the absence of pulses. After the correlation
at several shifted times (\texttt{TICK\_SHIFT}), characteristic values such as the
arrival time and pulse height (\texttt{PHA}; the amplitude of correlation) are derived
by interpolation. Here, the \texttt{TICK\_SHIFT} is a value that represents the extent
to which the template pulse is shifted in time relative to a incoming pulse from the
initial value. It is calculated by the PSP and takes integer values ranging from --8 to
$+$7 in the unit of a sample (80
$\mu$s)\cite{ishisakiInflightPerformancePulseprocessing2018,tsujimotoInOrbitPerformanceDigital2018}. For
L grade events, template fitting is not done and their pulses are characterized
similarly to anti-co events based on thresholding. Because of this difference, a
systematic offset of arrival time arises among different grades. This is calibrated as a
function of incoming x-ray energy for each channel
\cite{omamaRelativeTimingCalibration2022} and corrected on the ground to
$\lesssim$5~$\mu$s.

\subsection{Data processing on the ground}\label{s2-3}
The ground processing takes several stages from the level 0 to 3 (L0--L3). The L0
products are raw packet telemetry (RPT) files in the FITS format. The L1 products are
generated by the pre-pipeline (PPL) developed by
JAXA\cite{teradaDetailedDesignScience2021}, in which raw telemetry values are converted
to engineering values in the First FITS Files (FFF). The relative arrival time
correction is made at this stage. The L2 products are generated by the pipeline (PL)
developed by NASA\cite{doyleXRISMPipelineSoftware2022}, in which the engineering values in the L1 products are
converted into physical values based on the calibration data base (\texttt{CALDB}). The
L3 products are for quick-look assessment for scientific purposes such as x-ray images,
spectra, and light curves.

Event screening is performed in the PL using the \texttt{xapipeline} task for the
satellite-wide processing and the \texttt{rslpipeline} task for the Resolve-specific
processing. The following processing items are performed.
\begin{enumerate}
 \setlength{\itemsep}{-1mm}
 \item The good time intervals based on the South Atlantic Anomaly passages, Earth
       elevation angles, and others are calculated using the \texttt{xafilter} and
       \texttt{ahgtigen} tasks.
\item The time of the recycling operation of the adiabatic demagnetization refrigerator
      (ADR), the 50-mK cooler, is calculated using the \texttt{rslctsfluct} and
      \texttt{rsladrgti} tasks.
\item The time of modulated x-ray source (MXS) illumination is calculated using the
      \texttt{rslmxsgti} task.
\item Gain tracking and correction are performed for each event.
\item Various flags that can
      be used for potential later screening are assigned to each event using the
      \texttt{rslflagpix} and \texttt{rslplsclip} tasks.
\item Cleaned event files are generated for scientific usage using the \texttt{ahscreen} task.
\end{enumerate}

\section{Types of event screening}\label{sec:individual}
The Resolve event screening consists of 19 items in three categories as listed in
Table~\ref{tab:screeningitems}. The items in the first category are based on the pulse
shape (\S~\ref{sec:pulse_shape}). Characteristic values and flags are derived in the
onboard processing for each pulse, which are used to discriminate signal from background
events. Those in the second category are based on the relative arrival times of multiple
events arising from various physical processes that produce multiple events in
the microcalorimeter and/or the anti-co detector (\S~\ref{sec:timing}). Those in
the third category are based on the Good Time Intervals (GTI; \S~\ref{sec:gti}). Time
intervals not used for scientific purposes due to the spacecraft and instrument
operations are identified and removed.

\begin{table}[htpt!]
 \centering
 \caption{Resolve event screening items.}
 \begin{threeparttable}
  \begin{tabular}{lclcc}
   \hline
   Item & File\tnote{a} & Value\tnote{b} & Init scr\tnote{c} & Ref \\
   \hline
   \multicolumn{5}{c}{--- Screening based on event pulse shape ---}\\
   Rise time of a pulse & EVT & \texttt{RISE\_TIME} & No & \S~\ref{sec:pulse_shape} \\
   Time shift in the template fitting & EVT & \texttt{TICK\_SHIFT} & Yes & \S~\ref{sec:pulse_shape} \\
   Maximum of the time derivative of pulse & EVT & \texttt{DERIV\_MAX} & No & \S~\ref{sec:pulse_shape} \\
   Possible overlapping of pulses & EVT & \texttt{QUICK\_DOUBLE} & Yes & \S~\ref{sec:pulse_shape} \\
   Too slow rise or decay times & EVT & \texttt{SLOW\_PULSE}\tnote{d} & Yes & \S~\ref{sec:pulse_shape} \\
   Different slope in pulse decay & EVT & \texttt{SLOPE\_DIFFER} & Yes & \S~\ref{sec:pulse_shape} \\
   Possible saturation in the ADC & EVT & \texttt{FLAG\_CLIPPED} & No & \S~\ref{sec:pulse_shape} \\
   \hline
   \multicolumn{5}{c}{--- Screening based on relative event timing ---}\\
   Veto by the anti-coincidence detector & EVT & \texttt{STATUS[3]} & Yes & \S~\ref{s3-2-1} \\
   ``Short'' electrical cross-talk & EVT & \texttt{STATUS[7:8]} & No & \S~\ref{s3-2-2} \\
   ``Long'' electrical cross-talk  & EVT & \texttt{STATUS[13:14]} & No & \S~\ref{s3-2-3} \\
   Frame events & EVT & \texttt{STATUS[4]} & No & \S~\ref{s3-2-4} \\
   Electron recoil events & EVT & \texttt{STATUS[6]} & Yes & \S~\ref{s3-2-5} \\
   \hline
   \multicolumn{5}{c}{--- Screening based on time intervals ---}\\
   Recycling operation of the ADR & HK & \texttt{ADRC\_CT\_(CTL|MON)\_FLUC} & Yes & \S~\ref{s3-3-1} \\
   Calibration x-ray illumination (MXS) & EVT & \texttt{STATUS[9:12]} & Yes & \S~\ref{s3-3-2} \\
   Calibration x-ray illumination (FW $^{55}$Fe) & HK & \texttt{FW\_POSITION1/2} & Yes & \S~\ref{s3-3-2} \\
   \begin{tabular}[l]{@{}c@{}}
   Lost time due to the overload \\of the onboard CPU
   \end{tabular}
   & EVT & \texttt{STATUS[2]} & Yes & \S~\ref{s3-3-3} \\
   Passage through the South Atlantic Anomaly & EHK & \texttt{SAA\_SXS} & Yes & \S~\ref{s3-3-4} \\
   Occultation by the Earth & EHK & \texttt{ELV} & Yes & \S~\ref{s3-3-5} \\
   Cut-off rigidity & EHK & \texttt{CORTIME} & No & \S~\ref{s3-3-6} \\
   \hline
  \end{tabular}
 \label{tab:screeningitems}
 \begin{tablenotes}
  \item[a] Files where the screening information is available (EVT for event files, HK
  for house-keeping telemetry files, and EHK for the extended HK files).
  \item[b] Values used for screening.
  \item[c] Whether the item is included in the initial screening.
  \item[d] The most significant bit of \texttt{RISE\_TIME} is used for this flag.
\end{tablenotes}
 \end{threeparttable}
\end{table}

\subsection{Screening based on pulse shape}\label{sec:pulse_shape}
\begin{figure}[htpt!]
 \centering
 \includegraphics[width=1.0\columnwidth]{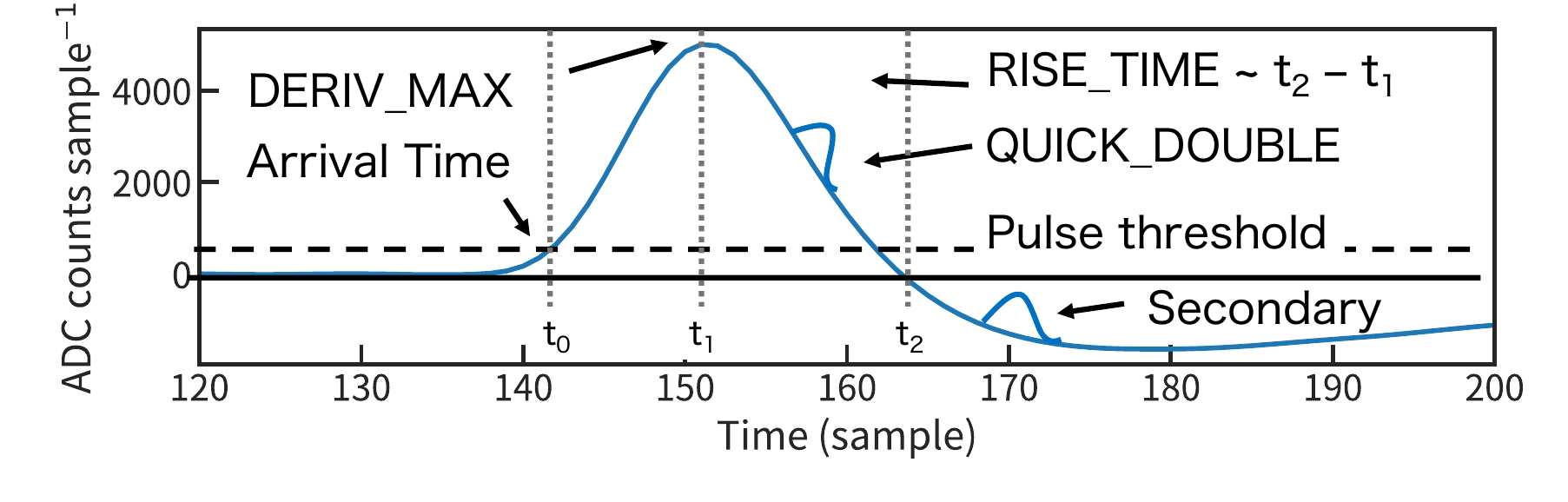}
 \caption{Conceptual example of the time derivative of an event.}
 \label{fig:pulse}
\end{figure}

The PSP derives several values and flags to characterize individual pulses
(Figure~\ref{fig:pulse}). \texttt{DERIV\_MAX} is the maximum value of the time
derivative. \texttt{RISE\_TIME} is defined as the time from the maximum time derivative
to the zero-crossing time interpolated for sub-sample resolution. Along with
\texttt{TICK\_SHIFT}, \texttt{PHA}, and arrival times (\S~\ref{s2-2}), they are
characteristic values of pulses. The maximum and minimum values of \texttt{TICK\_SHIFT}
are used to indicate that the template fitting was not successful. Events with such
values are screened. Other than this, the event properties are not used for the initial
screening.

\begin{figure}[htpt!]
 \centering
 \includegraphics[width=1.0\columnwidth]{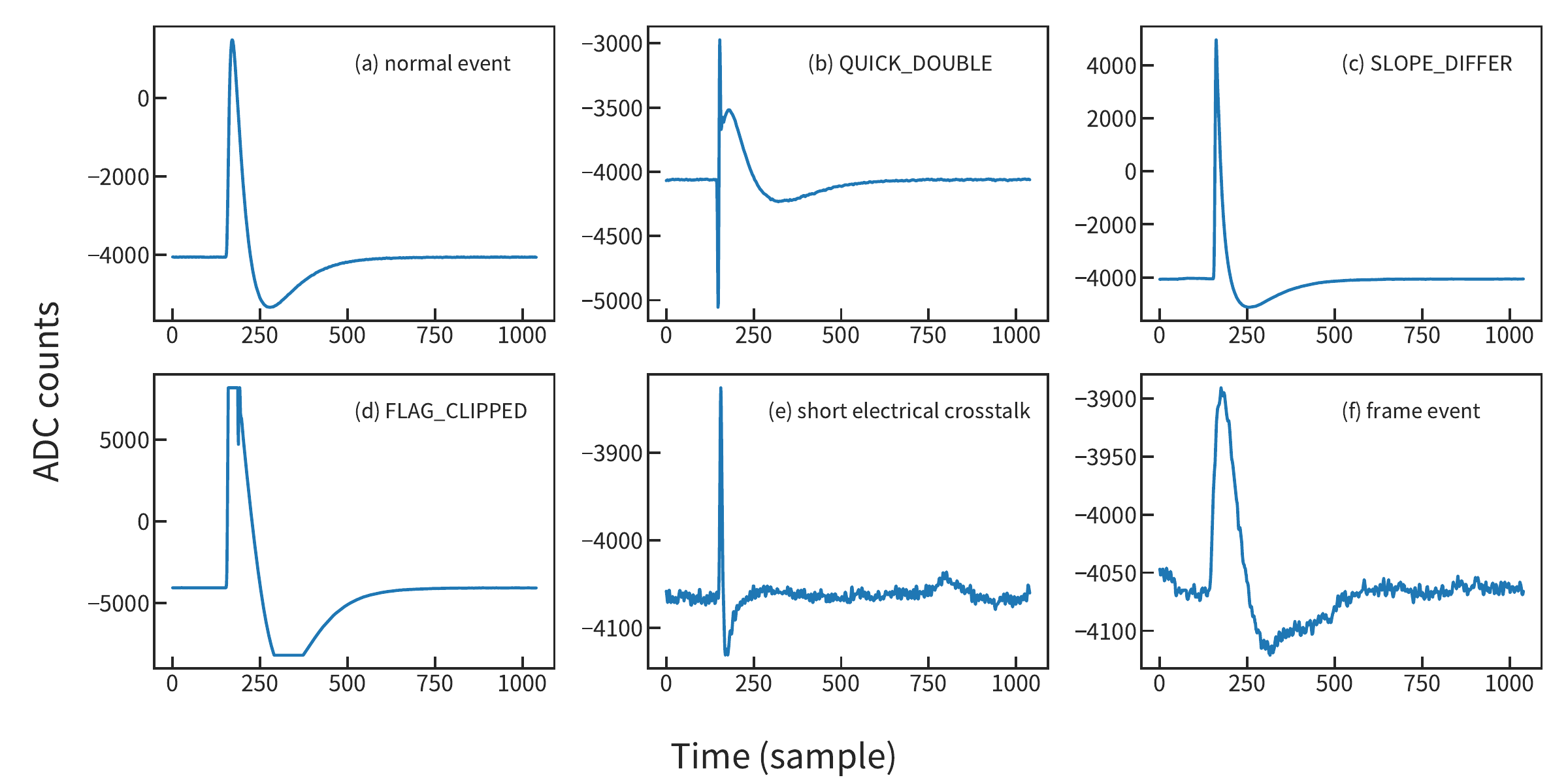}
 \caption{Some examples of the pulse record of (a) normal events and events with (b)
 \texttt{QUICK\_DOUBLE}, (c) \texttt{SLOPE\_DIFFER}, (d) \texttt{FLAG\_CLIPPED}, (e)
 short electrical crosstalk, and (f) frame event flags. Examples in (b)--(f) are not
 intended to be typical.}
 \label{fig:pulserecords}
\end{figure}

Flags are also given to indicate anomalies in the pulse shape
(Figure~\ref{fig:pulserecords}). The \texttt{QUICK\_DOUBLE} flag indicates an increase
of the time derivative during the decay phase before the trigger threshold is crossed
again, which corresponds to the rising edge of the original pulse.  \texttt{SLOW\_PULSE}
and \texttt{SLOPE\_DIFFER} flags are set for abnormal rise or slow time scales of
primary and secondary pulses, respectively\cite{ishisakiInflightPerformancePulseprocessing2018}.
Events with any of these flags are removed, except for events flagged with
\texttt{SLOPE\_DIFFER} that have $\texttt{PI} > 22000$. Many normal x-ray events at high
energy ($E$ $>$ 11~keV) receive the \texttt{SLOPE\_DIFFER} flag, and these events should not be removed.
We examined
pulse records of many events with \texttt{QUICK\_DOUBLE}, \texttt{SLOW\_PULSE}, and
\texttt{SLOPE\_DIFFER} flags found that they indeed exhibit abnormal pulse shapes.

\begin{figure}[htbp!]
  \centering
  \includegraphics[width=\columnwidth]{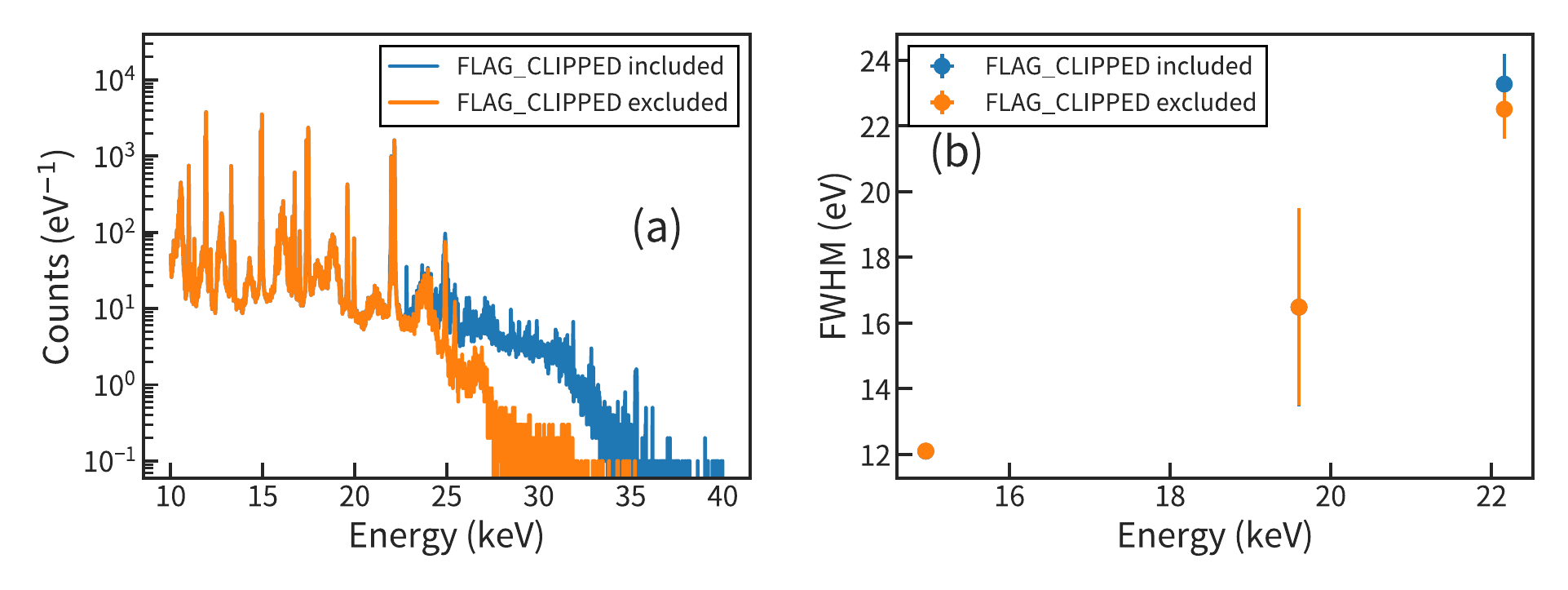}
 \caption{(a) Spectrum after initial screening with and without the
 \texttt{FLAG\_CLIPPED} flag. (b) The energy resolution with and without events of
 \texttt{FLAG\_CLIPPED} flag using the Y K$\alpha^{1}$ (15.0~keV), Mo K$\beta^{1}$
 (19.6~keV), and Ag K$\alpha^{1}$ (22.2~keV) lines. The results are almost identical for
 the first two lines.}
 \label{fig:clip}
\end{figure}

The \texttt{FLAG\_CLIPPED} flag is set for events with a possibility of hitting the
maximum of the ADC.
Such a flag could be given during the onboard processing when the entire pulse shape is available, but this function was not implemented.
In Resolve, the \texttt{rslplsclip} task was developed to flag
events of possible ADC saturation based on the pulse height\cite{eckart}. We examined
the algorithm using the ground data set taken from February 14, 2022, at 14:05 to
February 16, 2022, at 08:40 using a rotating target source, in which targets of an x-ray
generator rotate to provide multiple lines over a wide range of energy in the integrated
spectra in all event grades\cite{eckart2025}. Figure~\ref{fig:clip} shows the result,
in which most \texttt{FLAG\_CLIPPED} events appear at energies above 20~keV. Note that
the maximum energy before clipping depends on pixel and detector heatsink temperature.
Such events are found to degrade the energy resolution at 22 keV. The reason for this
change (for high and mid grade events) is because the shape of the gain scale is
significantly affected by the pulse clipping and not accounted for using the standard
gain correction techniques. This is not, however, included in the initial screening.
Many clipped events are background, from cosmic rays with long tracks in the absorbers,
but some may be signals, hence should be included with appropriate screening and
redistribution matrix functions in the future.


\subsection{Screening based on relative timing of multiple events}\label{sec:timing}
\subsubsection{Anti-coincidence window}\label{s3-2-1}
\begin{figure}[htpt!]
  \centering \includegraphics[width=\columnwidth]{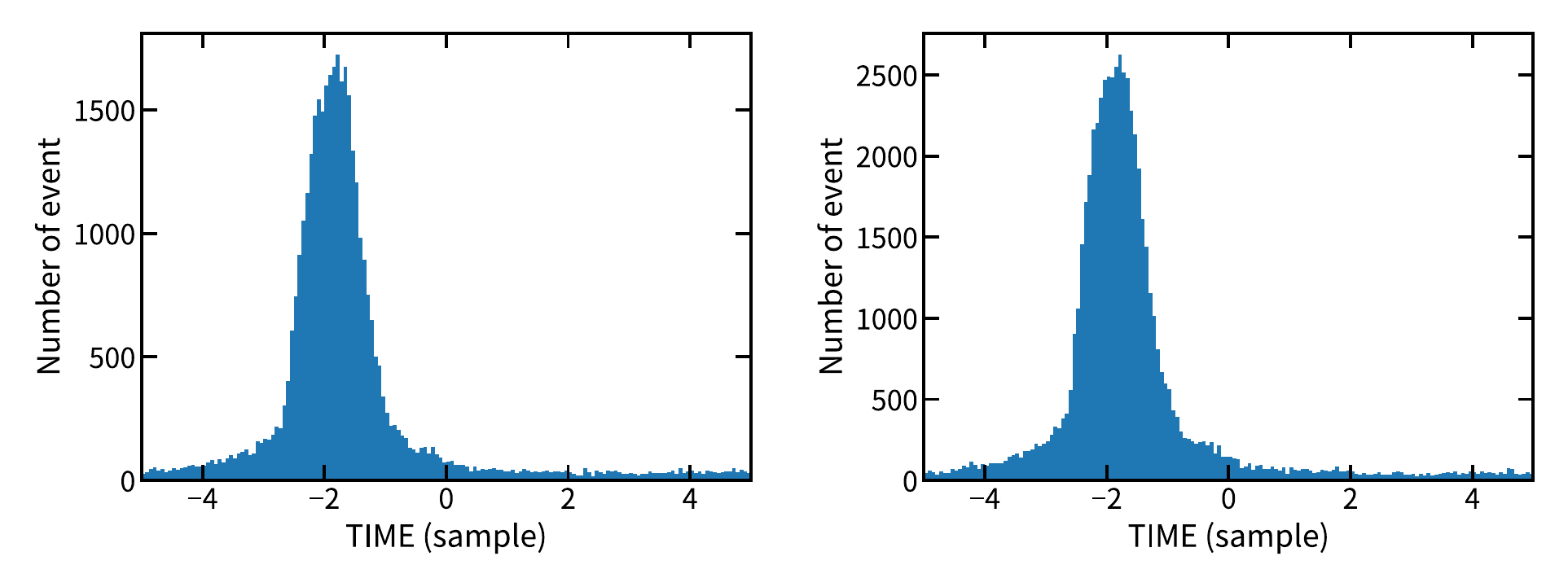}
 \caption{Arrival time of anti-co events relative to microcalorimeter events before
 (left) and after (right) the change of the event threshold on October 31, 2023. One sample time
 corresponds to 80~$\mu$s. The offsets for the different event thresholds agree with
 each other in the range of errors because the use of correlation with the template
 eliminates dependence of the arrival time on threshold.  different thresholds.}
 \label{fig:antico}
\end{figure}

We assessed whether the anti-coincidence window used for the SXS also works for Resolve
using its in-orbit data. Figure~\ref{fig:antico} shows the relative timing between the
anti-co and microcalorimeter events during times of no ADR recycle (\S~\ref{s3-3-1}), no
calibration x-ray illumination (\S~\ref{s3-3-2}), no SAA passages (\S~\ref{s3-3-4}), and
the night Earth elevation less than --5 degrees (\S~\ref{s3-3-5}). For the
microcalorimeter events, we applied the initial screening (\S~\ref{sec:initial}) except
for the anti-co screening, further applied the frame event screening
(\S~\ref{sec:timing}), and extracted Hp (high primary) grade (\S~\ref{s2-2})
events. The arrival time of anti-co events is earlier than the corresponding Hp
microcalorimeter events by --1.860 $\pm$ 0.004 and --1.859 $\pm$ 0.004 samples
respectively for the microcalorimeter event threshold of 75 and 120. The offset between
Hp microcalorimeter and anti-co events is caused by their different definition of
arrival times (\S~\ref{s2-2}). This is not corrected in the pipeline
processing. Instead, the anti-co window is currently set wide enough to encompass the
offset (6.25$\pm$6.25 samples). This is too wide for the distribution and a narrower
window (2$\pm$3 samples) would be sufficient.

\subsubsection{''Short'' electrical cross-talk}\label{s3-2-2}
Electrical cross-talk is an event for which a fraction of the signal power in a readout
line is transferred to the neighboring lines via capacitive coupling. This is considered
to happen in the high-impedance part of the readout in the cold stage between the
detector and the JFET. The pixel numbering of the microcalorimeter detector is based on
the line layout of this part, thus an event in pixel $i$ crosstalks to pixel $i-1$ and
$i+1$ except for boundaries. Anti-co channels also crosstalk to neighboring
microcalorimeter channels and vice versa. We call the original events parents and the
crosstalk events children hereafter.

\begin{table}[htpt!]
    \centering
 \caption{Data set used for the assessment of ``short'' cross-talk screening.}
 \begin{tabular}{ccccc}
  \hline
  Label & Start time & Stop time & Line & Energy (keV) \\
  \hline
  097091610 & 2022/02/01 04:00 & 2022/02/01 08:00 & $\mathrm{Fe~K} \alpha$ & 6.4 \\
  097091650 & 2022/02/01 20:00 & 2022/02/02 00:00 & $\mathrm{Au~L} \alpha$ & 9.7 \\
  097091750 & 2022/02/02 20:00 & 2022/02/03 00:00 & $\mathrm{Au~L} \beta$ & 11.4 \\
  \hline
 \end{tabular}
 \label{tab:cross_data}
\end{table}

Child events need to be removed as background, while parent events need to be left as
signals. ``Short'' crosstalk screening refers to the screening when the child events are
detected as events above the event threshold. ``Long'' crosstalk screening refers to
screening events contaminated by crosstalk that did not trigger but is inferred. For the
short cross-talk screening, we can utilize the relative timing and pulse height between
parent-child pairs in neighboring pixels. To better characterize the relation, we used
the data of monochromatic x-ray injection of a sufficiently high energy
\cite{leutenegger2020a} so that child events are detected. Three data sets of Fe
K$\alpha$ (6.4~keV), Au L$\alpha$ (9.7~keV), and Au L$\beta$ (11.4~keV) lines were used
(Table~\ref{tab:cross_data}). The Fe K$\alpha$ data represents the lowest parent energy
for which some child events can trigger for a threshold of 75. The Au L$\beta$
represents the highest end of the required energy band.


 \begin{figure}[htbp!]
  \centering
  \includegraphics[width=0.9\columnwidth]{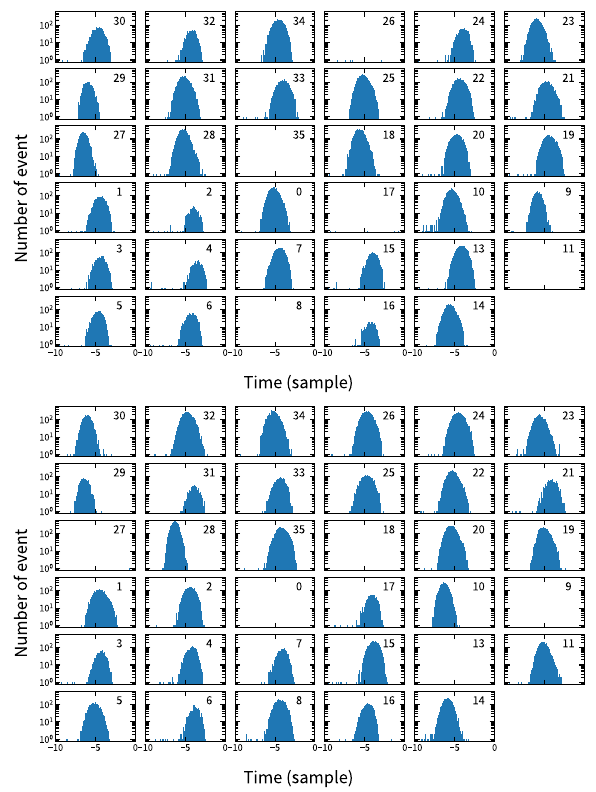}
 \caption{Result of short cross-talk assessment using the Au L$\alpha$ line data of
 individual pixels, arranged according to the physical layout of the pixels in the array, except for pixel 12 (calibration pixel). The distribution of arrival times of child
 events in pixel $i$ relative to their parents in pixel $i+1$ (top) and $i-1$
 (botom). The child events in pixels $9n+8 (n=0, 1, 2, 3)$ do not have parents in pixel
 $i+1$ and child events in pixels $9n$ do not have parents in pixel $i-1$ due to the
 boundaries of the readout line layout, which is confirmed with the data with a few
 false coincidence exceptions}.
 \label{fig:X-talk-6.43}
\end{figure}

\begin{figure}[htbp!]
  \centering
  \includegraphics[width=\columnwidth]{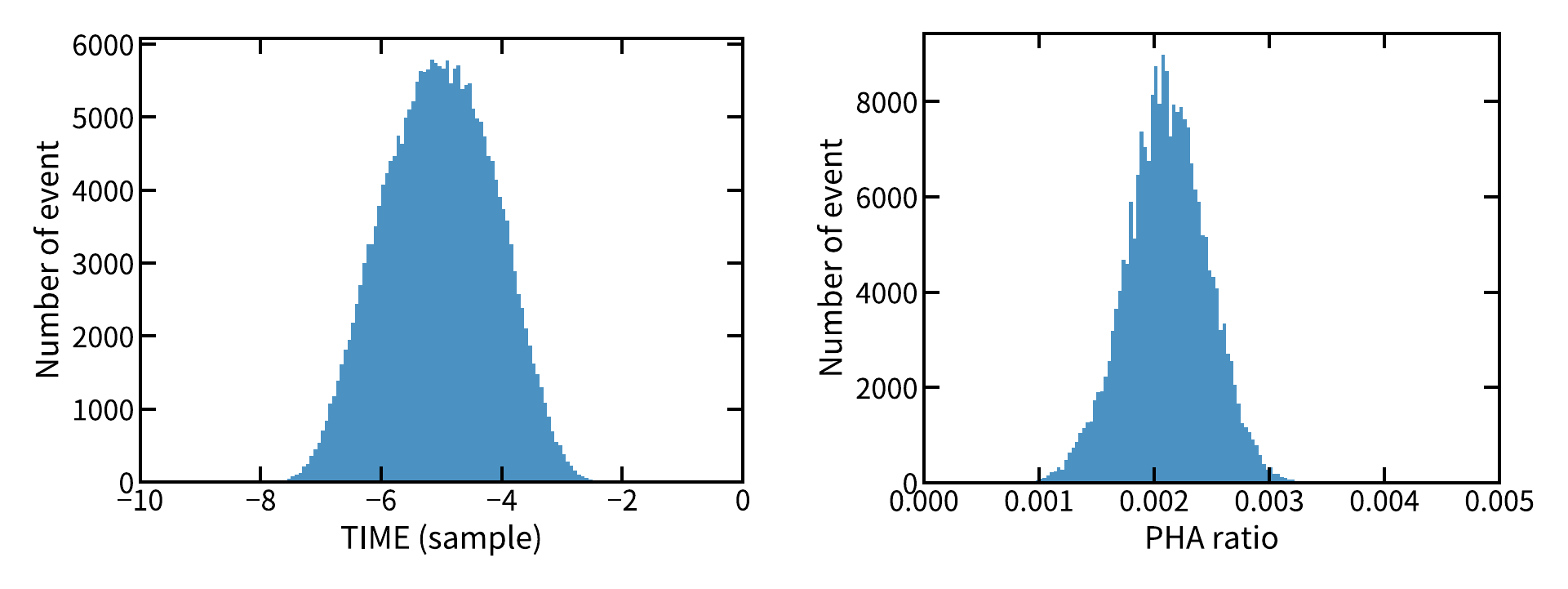}
 \caption{Result of short cross-talk assessment using the Au L$\alpha$ line data of all
 pixels except for pixel 12 (calibration pixel); (left) arrival time of child
 event candidates relative to their parents and (right) ratio of the pulse height of
 child events against parents. We used Hp grade only both for the parent and
 child events.}
 \label{fig:X-talk}
\end{figure}

Hp events in the energy range of $\pm$1~keV of the injected line energy were considered
parents. Child event candidates for each parent were extracted in two neighboring pixels
within a time window and having an energy less than
0.15~keV. Figure~\ref{fig:X-talk-6.43} and Figure~\ref{fig:X-talk} show the result using
the Au L$\alpha$ line as an example. Both the relative arrival time and the pulse height
ratio have characteristic distributions. For the former, child events precede their
parent events by $\sim$5 samples. This is reasonable as the child events are faster than
their parents due to their capacitive-coupling origin. For the latter, the ratio is
typically below 0.005. We made the same assessment for the other two data sets and
concluded that the relative window of 2--9 samples and the ratio below 0.005 encompasses
the parent-child pairs. This screening is effective for removing background below
$\sim$0.3~keV, which is relevant only after the GV is opened. Thus, the short crosstalk
flags \texttt{STATUS[7:8]} are assigned but not used as a part of the initial screening.

\subsubsection{``Long'' electrical cross-talk}\label{s3-2-3}

Long cross-talk screening is used to flag events in neighboring pixels that are likely
contaminated by crosstalk from each other, resulting in an error in the absolute energy
assigned to each pulse. The magnitude and polarity of the error in energy assignment
depends on how the crosstalk interacts with the template used to process the event and
thus on the offset of the pulses with respect to each other. It also depends on the
energy of the neighbor pulse and the coupling between the channels, which is not the
same for all neighboring pairs.  For bright sources and hard spectra, the accumulated
errors have a non-negligible effect compared with the required energy resolution. The
optimal screening is currently under development.  Careful trade-off studies depending
on science cases are needed.

Since the impact is causal, if the shapes of all the crosstalk pulses as a function of
energy were known, we could calculate the interaction with the template as a function of
offset and correct for it.  We may not need to know these pulse shapes very well to
significantly reduce the error, but more work is needed.

\subsubsection{Frame events}\label{s3-2-4}
Frame events are caused by cosmic rays hitting the detector frame made of silicon.  The
dissipated energy propagates as heat to the microcalorimeter pixels, which can detect
the disturbance as events if the trigger threshold is exceeded. These events are the
predominant background below $\sim$2 keV and should be discarded. Frame events exhibit
two distinctive characteristics because of their mechanism: (1) Multiple events are
detected within a short time period and (2) the detected events have a slow rise
time. Based on the first characteristics, we set the window to be $\pm$9 samples
\cite{kilbourneInflightCalibrationHitomi2018}. to flag the frame event candidates
(\texttt{STATUS[4]}). This screening, however, is based only on time coincidence, hence
is prone to false coincidence particularly in observations of bright sources. Therefore,
this is not currently included in the initial screening.

\subsubsection{Electron recoil}\label{s3-2-5}
An electron recoil event occurs when one of the electrons generated by an incoming x-ray
escapes from a pixel and deposits energy in a different pixel. Two events are recorded
and the original energy is reconstructed by summing the energy of the two. This
screening is possible only for pixel 12, for which the energy of x-ray photons is
limited mostly to 5.9~keV (Mn K$\alpha$) or 6.4~keV (Mn K$\beta$). Recoil events between
other pixels are removed by the same screening that removes the frame events.
The recoil events are distinguished by three criteria; (i)~They occur in pixels other
than pixel 12, (ii) The sum of the energies of the recoil event and the corresponding
event in pixel 12 is less than the energy of Mn K$\beta$ (6.5 keV, including margin),
and (iii) the difference of arrival times between the recoil and the original events is
within $\pm$3 samples. A flag is set for recoil events (\texttt{STATUS[6]}), which are
removed in the initial screening.

\subsection{Screening based on time intervals}\label{sec:gti}
\subsubsection{ADR recycle}\label{s3-3-1}
The working temperature for the heat sink of the microcalorimeter array is maintained at
50~mK using the ADR
\cite{shirron2025} and its controller, the ADRC.
Two ADRs are used in a series from the
1.2~K interface provided by the depressurized superliquid helium \cite{dipirro2025}. The magnet current
runs out in $\sim$2 days, and recycling takes $\sim$1~hour. Time
intervals during the recycling operation need to be removed.
The duration is based on the
control parameters and on the temperature fluctuations measured by a thermometer placed at the 50~mK
stage\cite{tsujimoto2018}. The algorithm was developed and verified anew for
Resolve, and is implemented in the \texttt{rsladrgti} task for the initial
screening. We confirmed that the new algorithm works by examining many ADR recycles
both on the ground and in orbit executed in various conditions.

\subsubsection{Calibration x-ray illumination}\label{s3-3-2}
Resolve has two types of calibration sources for tracking drifts in the energy
scales of individual pixels besides the $^{55}$Fe calibration source constantly
illuminating pixel 12 for tracking variations in common-mode gain\cite{porter2025}. One
is the modulated x-ray source (MXS)\cite{devries18,shipman2025,sawada2025}, in which
the calibration x-rays can be switched on and off at 1/8~ms at the shortest by
controlling the light emitting diode for the source of photoelectrons in the x-ray
generator\cite{sawadaPulseParametersOptimization2022}. Events during the illumination
intervals are distinguished by a set of flags (\texttt{STATUS[9:12]}). The other is the
$^{55}$Fe sources placed in one of the filter wheel (FW) positions\cite{devries18,shipman2025}. The wheel is rotated during observations to provide intermittent x-ray
illumination. Intervals of illumination are distinguished by the wheel position in the
HK telemetry (\texttt{FW\_POSITION}). Both the MXS and FW are provided by SRON. During
the closed GV configuration, we use the filter wheel $^{55}$Fe sources for the main
calibration source, as they can illuminate the entire array, unlike the MXS.

\subsubsection{Pixel lost time}\label{s3-3-3}
Due to the limited CPU resources onboard, all events are not processed at count rates
higher than $\sim$200~s$^{-1}$. In such a case, the buffer storing events detected by
the FPGA is discarded all together, and only the intervals of lost events are
downlinked. Each pixel has its own event buffer and each CPU processes nine pixels in a
round-robin manner. Therefore, the lost times occur differently among pixels. In most
cases, this happens when observing bright point-like sources and pixels at the array
center suffer the largest loss due to the CPU limit. The detailed assessment is
described separately\cite{mizumotoHighCountRate2025}.

\subsubsection{SAA passage}\label{s3-3-4}
XRISM passes through the South Atlantic Anomaly (SAA) 8--9 times lasting for
3--30~minutes each in a day of 15-16 rotations around the Earth. Resolve
continues to operate in observation mode during the SAA passages and events are
collected, which need to be removed based on the position of the spacecraft at the event
time.

\begin{figure}[htbp!]
  \centering
  \includegraphics[width=\columnwidth]{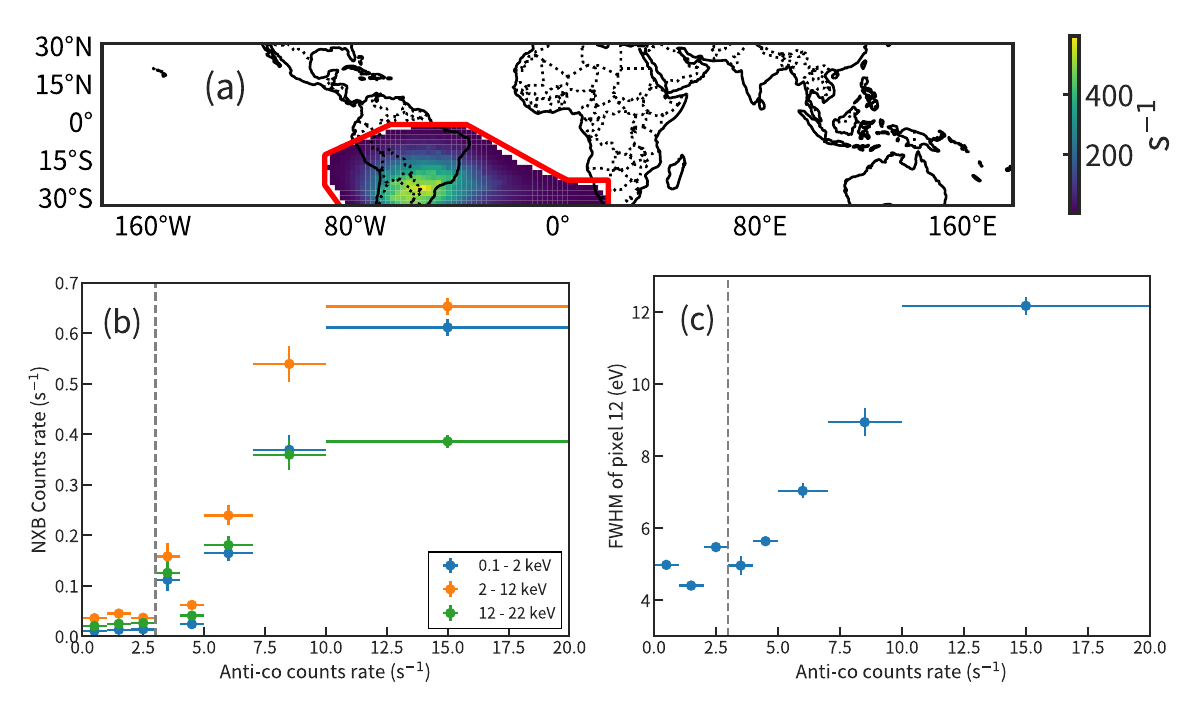}
 \caption{(a) Map of anti-co count rate averaged over a 2$\times$2 degree grid and red
 border defining SAA region (b) The NXB count rate (all event grades) of all pixels
 except for pixel 11, 12, and 13 (calibration pixel and its cross-talk neighbors). The
 rates in three different energy bands are shown in different colors. (c) The energy
 resolution in eV FWHM of the Mn K$\alpha$ line of pixel 12. Gray dash line in (b) and
 (c) shows the anti-co threshold used in the initial screening.}
 \label{fig:saa}
\end{figure}

As the SAA region changes gradually and has shifted westward significantly from the
ASTRO-H days, we redefined the region on the two-dimensional longitude--latitude
projection based on the in-orbit data of the anti-co detector. Figure~\ref{fig:saa}
shows the anti-co count rate in a 1~s bin averaged over a 2$\times$2 degree grid. We
used all the events taken from October 11, 2023, to December 6, 2023. We further removed
events in the time intervals of the ADR recycles (\S~\ref{s3-3-1}), x-ray illumination
(\S~\ref{s3-3-2}), and night earth elevation larger than --5
degrees. Figure~\ref{fig:saa} shows the non-x-ray background (NXB) count rate of the
array pixels and the energy resolution of pixel 12 as a function of the anti-co count
rate. As the count rate increases, the NXB rate increases and the resolution
degrades. We set the anti-co rate threshold to be $<$3~s$^{-1}$. We defined the SAA
region (red border in Figure~\ref{fig:saa}) so that all the grids exceeding the
threshold are enclosed.

\subsubsection{Earth occultation}\label{s3-3-5}
The time intervals when the observed source is occulted by the Earth are removed based
on the elevation angle of the source from the Earth limb. We inspected the background
rate of the in-orbit data and set the threshold to be $>$5 degrees for science
observations and $<$-5 degrees for NXB data collection, which are the same as used for
SXS. The threshold can be set differently for the Sun-lit Earth (day-Earth elevation)
and the Sun-unlit Earth (night-Earth elevation): the former may require a more stringent
criteria to avoid Earth albedo in the soft x-ray band. However, because of the closed GV
configuration, we see no difference between day and night Earth elevation, except when
large solar flares produce significant hard x-ray flux.  This will be revisited when the
GV is successfully opened.

\subsubsection{Magnetic cut-off rigidity}\label{s3-3-6}
The magnetic cut-off rigidity (COR) is a metric describing the strength of the Earth
magnetic field at the moving position of the satellite. When the COR is larger, the
cosmic-ray rate is reduced. The NXB rate and the energy resolution degradation depend on
COR only weakly\cite{kilbourneInflightCalibrationHitomi2018}, thus this metric is not
used for event screening purposes. Instead, the history of COR during an observation is
used to construct the background spectrum dedicated for the observation using the
\texttt{rslnxbgen} task. The verification of the task is underway as of this writing.

\section{Initial event screening}\label{sec:initial}
\begin{figure}[htbp!]
 \centering
 \includegraphics[width=1.0\textwidth, clip]{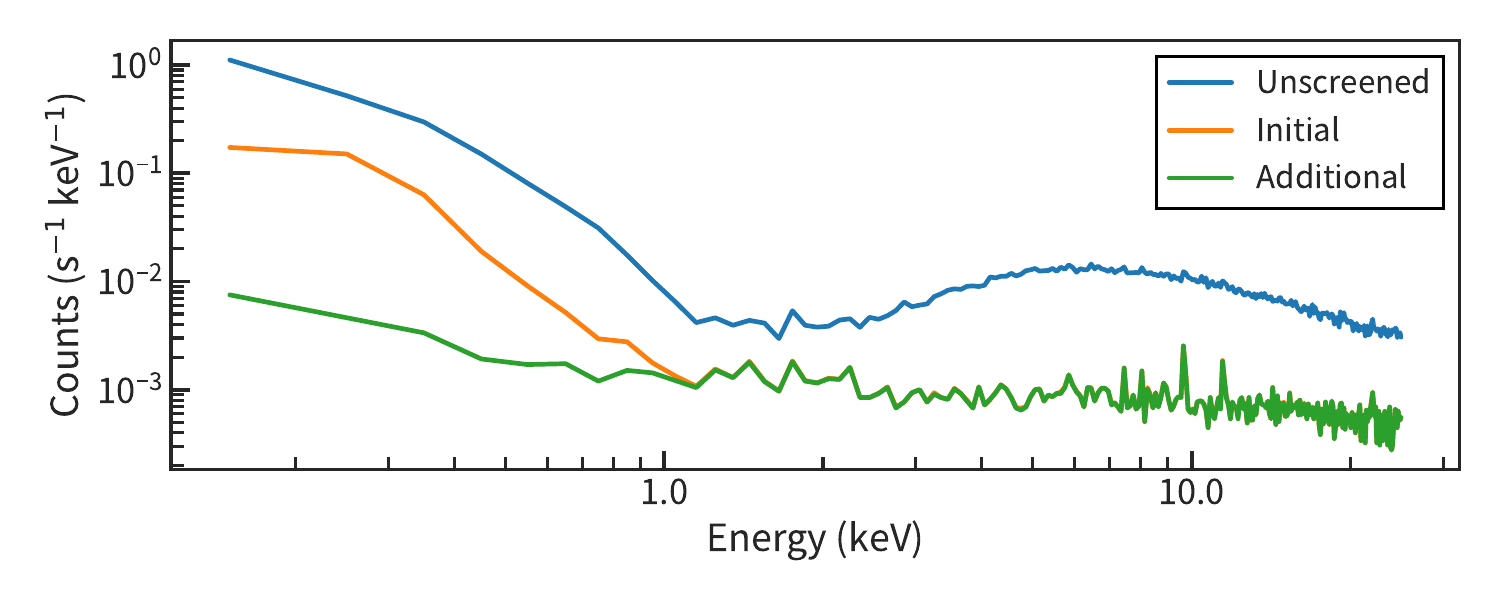}
 \caption{NXB spectra before any event screening (blue), initial screening (orange),
 and additional screening (green). Events of the Hp and Mp grade are used.}
 \label{fig:nxb}
\end{figure}

Here, the initial screening is the screening applied for the pipeline products for the
performance verification phase.  The screening can be applied to all events
independently of the observed source.  The screening is expected to be more
sophisticated as we gain more knowledge or conditions change in the future. A
combination of several screening items (Table~\ref{tab:screeningitems}) are used for the
initial screening. We now evaluate the performance of the initial screening using the
NXB data. Here, the NXB data are constructed using all the data from October 11, 2023
when the 50~mK control started to February 6, 2024 when the commissioning phase was
concluded. We merged data both before and after the change of the event threshold to
increase the statistics after confirming that the evaluation does not differ between the
two. The initial screening for NXB is applied (Table~\ref{tab:datasets}). Time intervals
of x-ray illumination were also removed. A total exposure time of 650~ks was
obtained. The resultant NXB spectrum is shown in Figure~\ref{fig:nxb}. The count rate in
the 0.3--12~keV band is 1.8$\times10^{-3}$~s$^{-1}$~keV$^{-1}$. We conclude that the
initial screening is sufficient to meet the requirement of
$<$2.0$\times10^{-3}$~s$^{-1}$~keV$^{-1}$.

\section{Additional event screening}\label{sec:additional}
\subsection{Screening criteria}\label{sec:2D screening}
Although the initial screening in the current pipeline processing is sufficient to
satisfy the requirement (\S~\ref{sec:initial}), additional screening, such as applied to
the ASTRO-H SXS data, but optimized for Resolve, is expected to improve the
signal-to-noise ratio and extend the Resolve capability for specific sources.
We exploit the fact that the characteristic values, not flags, of events (i.e.;
\texttt{DERIV\_MAX}, \texttt{RISE\_TIME}, and \texttt{TICK\_SHIFT}) are tightly
correlated with each other for x-ray events but not for background. For example, the
higher the energy deposited, the greater the drop in the resistance of the thermistor,
and the resulting change in the electrical component of the signal response time leads
to a correlation between \texttt{DERIV\_MAX} and \texttt{RISE\_TIME}. Events deviating
from this correlation should be considered as background events. Here, we used
\texttt{DERIV\_MAX} as opposed to another proxy of energy used in the previous
work\cite{kilbourneInflightCalibrationHitomi2018}, as the correlation is more linear
between \texttt{DERIV\_MAX} and \texttt{RISE\_TIME}. Other criteria such as frame
(\S~\ref{s3-2-4}) and cross-talk (\S~\ref{s3-2-2}, \ref{s3-2-3}) events are expected to
improve screening further, but they are not presented in this work as they are currently
under study.

\begin{figure}[htbp!]
  \centering
  \includegraphics[width=\columnwidth]{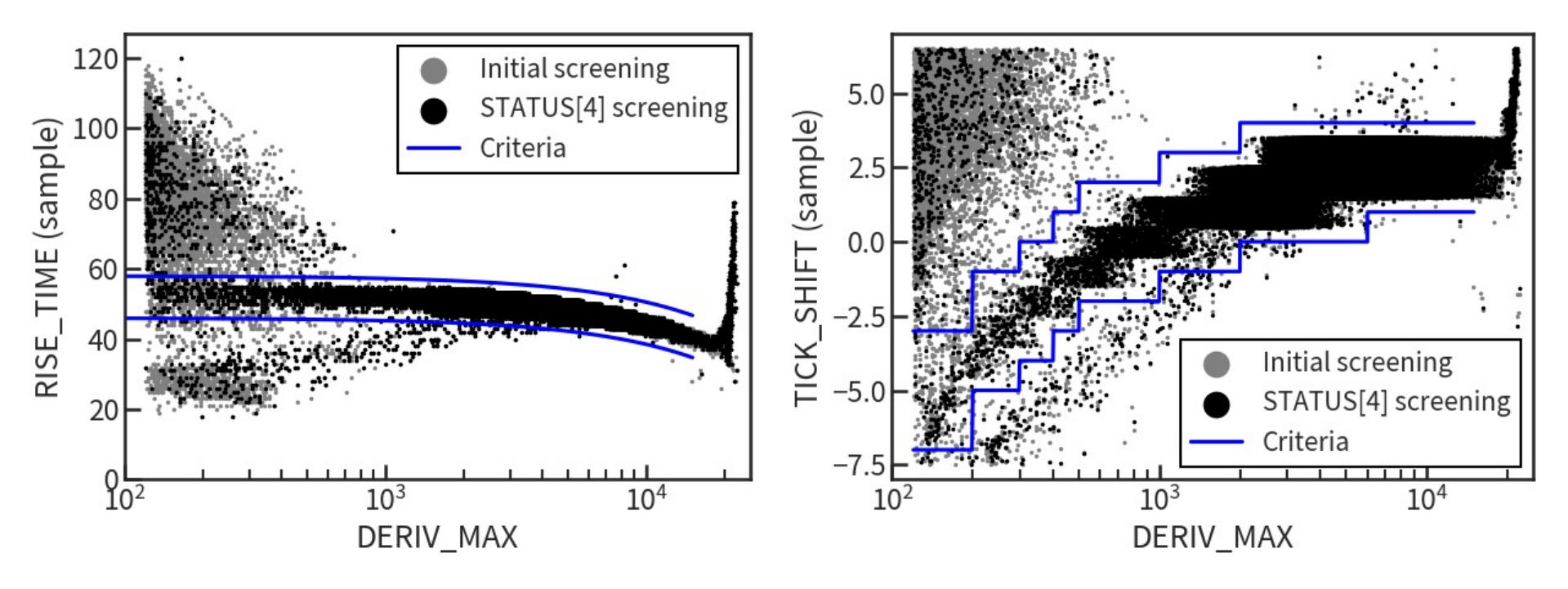}
 \caption{(left) \texttt{RISE\_TIME} versus \texttt{DERIV\_MAX} and (right)
 \texttt{TICK\_SHIFT} versus \texttt{DERIV\_MAX} distribution of Hp and Mp events using
 the in-orbit data of LMC X-3. The grey dots are events after the initial screening. The
 black dots are those after the additional screening to remove frame events. The value
 of \texttt{TICK\_SHIFT} is dithered by $\pm$0.5 for presentation purposes. With the GV
 closed, events with an energy below $\sim$1.7~keV cannot have originated from LMC X-3,
 but we expect real signal events to follow the trend shown here.}
 \label{fig:2D}
\end{figure}

Figure~\ref{fig:2D} shows the two-dimensional distribution of Hp and Mp events using the
in-orbit data of LMC X-3 observed in late November and early December, 2023. The gray
dots indicate events after the initial screening. The correlations among the x-ray
events are evident as the main single sequence. The distributions deviating from the
main sequence are mostly made with background events, which is confirmed by inspecting
the pulse shapes when the pulse record data are available. An exception is the
distribution deviating upward from the main sequence at the largest \texttt{DERIV\_MAX}
end, which is made by x-ray events that are clipped.
These events may be signals,
hence should not be removed. Only for the purpose of clarifying the main sequence, we applied
the frame-event screening (\S~\ref{s3-2-4}), which efficiently eliminates many
background events though with false coincidence. Based on the frame-screened distribution
of events (black dots), we defined the screening criteria shown in blue lines.
Note that the presented screening applies for the data taken with the event threshold of
120.
\texttt{TICK\_SHIFT} depends on the trigger threshold, thus different criteria for
\texttt{TICK\_SHIFT} were developed for the data taken with the event threshold of 75.

\subsection{Evaluation}\label{sec:add-evaluation}
We now evaluate the performance of the additional screening. The result of screening
depends on the count rate of the celestial object; the background may be reduced further
but at a sacrifice of data due to false coincidence. There may be some pixel
dependence. For the assessment, therefore, we used the data sets of different origins
(Table~\ref{tab:datasets}), which include the ground data with the GV open (GVO) and the
in-orbit data of Abell 2319 and LMC X-3 observations and NXB data.

\begin{table}[htbp!]
 \centering
 \begin{threeparttable}
 \caption{Data sets used for the assessment of additional screening.}
 \begin{tabular}{lccccc}
  \hline
  Label                    & Start time       & Stop time        & Exposure (ks) & Count rate$^{a}$         & Event threshold\\ \hline
  GVO                      & 2022/03/04 01:30 & 2022/03/09 23:00 & 403   & 3.2               & 75\\
  \hline
  Abell 2319               & 2023/10/14 07:25 & 2023/10/23 20:30 & 193   &  0.46             & 75\\
  \hline
  \multirow{2}{*}{LMC X-3} & 2023/11/16 12:05 & 2023/11/17 13:50 & \multirow{2}{*}{204} &  \multirow{2}{*}{0.49} & \multirow{2}{*}{120}\\
                           & 2023/11/30 08:04 & 2023/12/03 06:23 &                      & \\
  \hline
 \multirow{2}{*}{NXB} & 2023/10/11 23:13 & 2023/10/31 19:50 & \multirow{2}{*}{650} & \multirow{2}{*}{0.43} & 75 \\
                        & 2023/10/31 19:50 & 2024/01/04 15:07 &        & &       120 \\
  \hline
 \end{tabular}
 \label{tab:datasets}
         \begin{tablenotes}
            \item[a]  Counts rate of Hp and Mp grade events in 0.1--12~keV (s$^{-1}$ pixel$^{-1}$)
        \end{tablenotes}
  \end{threeparttable}
\end{table}

Abell 2319 is a galaxy cluster and is a diffuse source with a uniform distribution of
events over the microcalorimeter array\cite{feretti1997}. LMC X-3 is a blackhole binary
in the Large Magellanic Cloud\cite{cowley1983} and a point source making a concentrated
event distribution at the array center. We conducted spectral fitting of LMC X-3, the
brightest of the two, and found that the X-ray signals below 1.7~keV is negligible, thus
we use this energy as the boundary. The GVO data set was taken when the GV was open on
the ground and provides data below $\sim$1.7~keV, a region probed only by background
events and spectral redistribution when the GV is closed.

We apply the additional screening and see how the background and signal events are
affected. For signal events, we focus on the Hp and Mp grade events used for
spectroscopy. Signal and background events dominate different energy bands for these
data sets. Signal events dominate the whole band (0.1--20~keV) for the GVO data and in
the science band (1.7--12~keV) for the two observations in the closed GV
configuration. Background events dominate the whole band for the NXB data and in the
soft band $<$1.7~keV for the two observations.

\begin{table}[htbp!]
 \centering
 \begin{threeparttable}
  \caption{Ratio of the event rate after the additional screening to that before
  separately for the signal- or noise-dominant energy bands.}
  \begin{tabular}{ccc}
   \hline
   Label                    & Energy band (keV)    & Ratio$^a$ \\
   \hline
   \multicolumn{3}{c}{Signal-dominant band}\\
   \hline
   GVO         & 1.7--12 & 0.998 \\
   Abell~2319  & 1.7--12 & 0.95 \\
   LMC X-3     & 1.7--12 & 0.98 \\
   \hline
   \multicolumn{3}{c}{Noise-dominant band}\\
   \hline
   Abell~2319 & 0.1--1.7 & 0.05 \\
   LMC X-3    & 0.1--1.7 & 0.22 \\
   NXB        & 0.1--1.7 & 0.05 \\
   \hline
  \end{tabular}
  \label{tab:evaluating}
  \begin{tablenotes}
   \item[a] Ratio of the event rates after the additional screening to that before.
  \end{tablenotes}
 \end{threeparttable}
\end{table}

Table~\ref{tab:evaluating} shows the ratio of the event rate after screening to that
before in the energy band where signal or noise events dominate.  The reduced rate in
the signal-dominant band is due either to the false coincidence or contamination of
background events in the band.
%
More events are removed by screening for the lowest count rate data set of Abell 2319
than for the other two sets. This is likely because background events are non-negligible
in the signal-dominant band of the lowest count rate data set. Based on these data, we
conclude that the percentage of real source data removed by the additional screening is
less than 6\%.
For Abell 2319 and the NXB, the additional screening reduced the rate by $\sim$95\%
below 1.7~keV. The reduction is lower for the highest count rate data set of LMC
X-3. This is likely due to spectral redistribution from the signal dominant band at
$>$1.7~keV to the background dominant band at $<$1.7~keV\cite{tsujimoto2018}. These
redistributed events, resulting from phenomena such as the removal of energy by escaping
photons or electrons after absorption of an x-ray, represent signal information, hence
should not be removed as background but should be modeled through the redistribution
matrix.

The resultant NXB spectrum after the additional screening is shown in
Figure~\ref{fig:nxb}. The count rate is $1.0 \times 10^{-3}$~counts~s$^{-1}$~keV$^{-1}$
in the 0.3--12~keV band, which is smaller than the requirement by a factor of 2. The
additional screening is particularly useful below $\lesssim$2~keV, which will be
accessible after the GV opens in the orbit.

\section{Conclusion}\label{sec:conclusion}
We presented the result of Resolve event screening using data acquired on the ground and
in orbit. We assessed and optimized the screening criteria of individual screening items
in three categories. New screening items (relative to SXS screening) for clipped events
and ADR recycles were validated. More optimal screening criteria for the
anti-coincidence detector, cross-talk events, and the SAA were developed. With the
initial screening, we confirmed that the residual background rate is $1.8 \times
10^{-3}$~s$^{-1}$~keV$^{-1}$ in the 0.3--12.0~keV range to meet the requirement. With
the additional screening based on the relationship between the pulse-shape properties
characteristic of x-ray events, we found that the background is reduced further, in
particular in the $\lesssim$2~keV band, and becomes $1.0 \times
10^{-3}$~s$^{-1}$~keV$^{-1}$ in 0.3--12~keV.
The additional screening will be particularly useful, when the GV is opened in the future, for soft diffuse extended sources such as supernova remnants. We will work to extend the bandpass beyond the requirement to allow access to the carbon features on the lower end and the Compton hump of many astrophysical objects on the high end.  We expect refinements to the screening to be necessary.  Currently, the multiple coincidence-based screening criteria overlap, and we are in the process of adding additional criteria to each, such as minimum and maximum ratios and size of the coincident group, such that the screening steps will be as distinct and precise as possible.

\acknowledgments

The substantial contributions of every member of the XRISM Resolve team enabled this
work. We appreciate careful reading and useful comments for improvements by two
anonymous reviewers.  This work was supported by JSPS Core-to-Core Program (grant
number:JPJSCCA20220002).  This research has made use of data and software provided by
the High Energy Astrophysics Science Archive Research Center (HEASARC), which is a
service of the Astrophysics Science Division at NASA/GSFC.  This work was supported by
JST SPRING, Grant Number JPMJSP2108 and by Iwadare Scholarship Foundation in 2023.  Part
of this work was performed under the auspices of the U.S. Department of Energy by
Lawrence Livermore National Laboratory under Contract DE-AC52-07NA27344.  This paper is
based on the SPIE proceeding\cite{mochizuki2024}.

\section*{DATA AVAILABILITY}
The data that support the findings of this article are proprietary and are not publicly available.
The data displayed in the figures above, as well as a limited subset of the underlying data, are available upon request at \linkable{mochizuki@ac.jaxa.jp}.

\section*{Disclosures}
The authors declare there are no financial interests,
commercial affiliations, or other potential conflicts of interest that have influenced the objectivity of this research or the writing of this paper.


\bibliography{main}   
\bibliographystyle{spiejour}   


\vspace{2ex}\noindent\textbf{Yuto Mochizuki} is a graduate school student of The
University of Tokyo and Institute of Space and Astronautical Science, Japan Aerospace
Exploration Agency. He received his BS degree in physics from Tokyo University of
Science and MS degree in astronomy from The University of Tokyo in 2022 and 2024,
respectively.  He is a student member of SPIE.

\vspace{1ex}
\noindent Biographies and photographs of the other authors are not available.

\listoffigures
\listoftables

\end{spacing}
\end{document}